# THE HEAT CAPACITY OF NITROGEN CHAINS IN GROOVES OF SINGLE-WALLED CARBON NANOTUBE BUNDLES

**M.I. Bagatskii, M.S. Barabashko, V.V. Sumarokov**

*B. Verkin Institute for Low Temperature Physics and Engineering, 47 Lenin Ave., 61103 Kharkov, Ukraine*

e-mail: bagatskii@ilt.kharkov.ua

**Abstract**

The heat capacity of bundles of closed-cap single-walled carbon nanotubes with one-dimensional chains of nitrogen molecules adsorbed in the grooves has been first experimentally studied at temperatures from 2 K to 40 K using an adiabatic calorimeter. The contribution of nitrogen $C_{N2}$ to the total heat capacity has been separated. In the region $2 - 8$ K the behavior of the curve $C_{N2}(T)$ is qualitatively similar to the theoretical prediction of the phonon heat capacity of 1D chains of Kr atoms localized in the grooves of SWNT bundles. Below 3 K the dependence $C_{N2}(T)$ is linear. Above 8 K the dependence $C_{N2}(T)$ becomes steeper in comparison with the case of Kr atoms. This behavior of the heat capacity $C_{N2}(T)$ is due to the contribution of the rotational degrees of freedom of the $N_2$ molecules.

# 1. Introduction

Since the discovery of carbon nanotubes in 1991 [1], investigations of the physical properties of these novel materials have been rated as a fundamentally important trend in physics of condensed matter [2,3]. The immense practical and scientific interest in carbon nanostructures stems from their unique physical (mechanical, electrical, magnetic, optical and so on) [3-19] characteristics.

Carbon nanotubes have a large specific surface and are promising as adsorbents for solving several technical problems such as storage of gaseous and condensed substances or separation of isotopes and gas mixtures [20,21].

The adsorption of gases and the physical properties of gases adsorbed by bundles of closed carbon nanotubes (c-SWNT) are of fundamental interest in the physics of low-dimensional systems [22-27]. The structure of c-SWNT bundles enables formation of 1-, 2- and 3-dimensional systems.

Technologically, most of the tubes in as-prepared bundles have closed ends unless special steps are taken to open them up. Owing to Van der Waals attractive forces, nanotubes can unite into bundles. Within a bundle the nanotubes form a close-packed two-dimensional (2D) triangular lattice [28].

The possible sites of adsorption of gas impurities in c-SWNT bundles are interstitial channels (IC), grooves (G) and the outer surface (OS) (see Fig. 1). These sites differ in geometric size and binding energy [29-32]. At low adsorbate concentrations, one-dimensional (1D) chains of impurity molecules (atoms) are formed in the IC- and G-sites. One or several layers of molecules (atoms) adsorbed at the outer surface of the c-SWNT bundle form quasi-two-dimensional (2D) or quasi-three-dimensional (3D) systems. The 1D, 2D and 3D systems have different properties at low temperatures [6,33-39].

The physical adsorption of gases by c-SWNT bundles and the adsorbate dynamics have been the subject of numerous theoretical and experimental investigations. The adsorption of nitrogen by c-SWNT bundles was considered in [14,31,40-42]. The low-temperature thermodynamics of helium adsorbed in the grooves of c-SWNT bundles was analyzed theoretically using the lattice gas model [27,43]. The quantum states and the heat capacity of low-density $^4$He gas adsorbed in the interstitial channels of c-SWNT bundles were studied theoretically in Refs. [36,44,45]. The heat capacity of c-SWNT bundles with adsorbed Ne and Xe atoms was investigated by the Monte Carlo method [46]. The phonon heat capacity of 1D chains of inert gases atoms (Xe, Kr, Ar, Ne) and $CH_4$ molecules adsorbed in the grooves at the outer surface of c-SWNT bundles, was calculated in Refs. 47,48. The authors of article [48] also calculated the phonon heat capacity at a constant volume for three chains of $CH_4$ molecules adsorbed in the grooves.

Experimental investigations of the thermal properties of c-SWNT bundles containing adsorbed gases are only at the initial stage. For the most part the results obtained refer to the coefficients of the radial thermal expansion $\alpha_r$ of c-SWNT bundles saturated with gases ($^3$He, $^4$He, H$_2$, N$_2$, O$_2$, Xe) [49-55] above 2 K. It is found that in the investigated temperature intervals $α_r$ increases sharply in c-SWNT bundles saturated with H$_2$, N$_2$, O$_2$, Xe [49-51,53] and the dependences $α_r(T)$ exhibit maxima. It was assumed that the maxima in the dependences $α_r(T)$ account for the spatial redistribution of adsorbate particles at the surface of c-SWNT bundles. The saturation of c-SWNT bundles with helium isotopes ($^4$He [52] and $^3$He [54]) causes a dramatic increase in the magnitude of the negative thermal expansion in the interval 2.1–7 K. It is believed [52,54] that the effect can be induced by the tunneling motion of helium atoms during of their spatial redistribution. The detected great isotope effect is due to the fact that $^3$He atoms have a smaller mass than $^4$He and hence a higher probability of tunneling [19,54].

The heat capacity of $^4$He-saturated SWNT bundles was investigated below 6 K [56,57]. The heat capacity of the adsorbed $^4$He exhibited a quasi-two-dimensional behavior ($C_{ads} \sim T^2$) in the sample of SWNT bundles prepared by laser vaporization technique and a quasi-one-dimensional behavior ($C_{ads} \sim T$) in the sample prepared by the arc discharge method [57].

Recently, precise measurements of heat capacities have been performed for the first time on c-SWNT bundles with adsorbed 1D chains of Xe atoms in the grooves at temperature range from 2 to 30 K [58,59]. The experimental heat capacity agrees within the measurement error with the theoretical curve [47] at $T < 8$ K.

This study continues calorimetric experiments carried out in the same adiabatic calorimeter and on the same sample of c-SWNT bundles saturated with different gases. The heat capacity of bundles of closed-cap single-walled carbon nanotubes with one-dimensional chains of nitrogen molecules adsorbed in the grooves has been first experimentally studied at temperatures from 2 K to 40 K using an adiabatic calorimeter. Behavior of simple molecular adsorbates, such as N$_2$ under similar conditions can be interesting for a few reasons. First, will quasi-one-dimensional N$_2$ chains demonstrate alike thermodynamic properties and, second, are one to expect certain peculiarities related to fact that the dopant particle is a molecule. In addition, the nitrogen molecule has a quadrupole moment, which can enrich the emerging picture.

**2. Experiment**

The adiabatic calorimeter and the measurement technique have been reported in Ref.. 60. The temperature in the calorimeter was measured with a calibrated CERNOX resistance thermometer (Lake Shore Cryotronics). Precise measurements of the heat capacity of the calorimetric vessel with a sample of pure c-SWNT bundles ("addenda") were made in Ref. 6.

A cylindrical sample of c-SWNT bundles (7.2 mm high, 10 mm in diameter, of 1.27 g/cm$^3$ density) was prepared by compressing c-SWNT plates under the pressure 1.1 GPa. The plates (~0.4 mm thick) were obtained by compacting a SWNT powder ("Cheap Tubes") also under the pressure 1.1 GPa. The powder was prepared by chemical catalytic vapor deposition (CVD). It contained over 90 wt % of SWNT bundles, other allotropic forms of carbon (fullerite, multiwalled nanotubes and amorphous carbon) and about 2.9 wt% of cobalt catalyst. The average tube diameter in the sample was 1.1 nm, the average length of the SWNT bundles was 15 μm. The number of nanotubes in the bundles varied within 100 – 150 (estimated from high – resolution TEM pictures). The mass of the sample of c - SWNT bundles was 716.00±0.05 mg [6].

This experiment was made after measuring the heat capacity of c-SWNT bundles with 1D chains of Xe atoms in the grooves at the outer surface of the bundles [58,59]. Before starting the experiment, the vacuum chamber of the calorimeter with a c-SWNT bundle sample at room temperature was washed several times with pure $N_2$ gas and the sample stayed in dynamic vacuum (~5·$10^{-3}$ Torr.) for about 12 hours. Then test measurement of the "addenda" heat capacity was performed. The results coincided with the data of Ref. 6. After completing the "addenda" measurement, the vacuum chamber of the calorimeter was filled with nitrogen at room temperature. The quantity of nitrogen was found by the PVT method ($\mu_{N2}$= 3.81·$10^{-4}$ ± 6·$10^{-6}$ mol). The chemical purity of $N_2$ was 99.997% ($O_2 \leq$ 0.003%). The sample of c-SWNT bundles was saturated with $N_2$ directly in the vacuum chamber of the calorimeter by cooling the calorimeter cell. The ratio of the number of $N_{N2}$ molecules adsorbed by the c-SWNT bundles to the number of carbon atoms $N_C$ in the sample of c-SWNT bundles was estimated to be $\xi_{N2} = N_{N2}/N_C \approx 0.0066$. The parameter $\xi_{N2}$ characterizes the occupancy of c-SWNT bundles with nitrogen.

The quantity of $N_2$ required to form one chain in all grooves of the c-SWNT bundles was estimated in advance. The calculation using a geometric model and assuming that the average tube diameter was 11Å and the average number of tubes in a bundle was 127. We also assumed that the distance between the $N_2$ molecules in the 1D chains was $a$ = 3.994Å, which corresponds to the nearest neighbor distance in the $Pa$3 lattice of solid $N_2$ at $T$ = 0 K [61].

On cooling the calorimeter cell the $N_2$ molecules are adsorbed first of all in the grooves because at these sites their binding energy is higher than at the outer surface of the bundles [31]. $N_2$ molecules are unable to penetrate into the interstitial channels because their cross-section sizes are smaller than the size of $N_2$ molecules [29,62]. During $N_2$ adsorption the temperature distribution in the vacuum chamber of the calorimeter together with calorimetric cell must be maintained at the level permitting the c-SWNT bundles to adsorb all the nitrogen available in the vacuum chamber.

After filling the vacuum chamber of the calorimeter with $N_2$ ($\mu_{N2}$= 0.000381 mol) at room temperature the pressure of the $N_2$ gas in the chamber was ~ 16 Torr. As follows from the equilibrium vapor pressure above solid nitrogen [63], this value corresponds to $T$~56 K of solid $N_2$.

Since the moment of filling the $N_2$ bath of the cryostat with liquid $N_2$, the calorimeter cooled down from 289 K to 90 K during ~8 hours. Then, the temperature $T_{VC}$ of the vacuum chamber walls was decreased by blowing cold $^4$He gas through the helium bath of the cryostat. To increase the "effective" enthalpy of the $^4$He gas in the helium bath, before the experiment the lower part of the helium bath was filled with adsorbent $Al_2O_3$ (500 cm$^3$). This permitted us to reduce the derivative $dT_{VC}/dt$ almost by an order of magnitude in the process of blowing cold $^4$He gas through the helium bath of the cryostat and cool the calorimeter from 90 K to ~60 K during ~5 hours. Then the helium bath of cryostat was filled with liquid $^4$He.

According to the isotherm of $N_2$ adsorption in samples of SWNT bundles at $T$ = 77 K [31], saturation to $\xi_{N2}$ ≈0,0066 produces the pressure ~$10^{-4}$ Torr of the $N_{N2}$ gas over the sorbent. In the case of nitrogen adsorption isotherm at 60 K the nitrogen gas pressure is ~ $10^{-5}$ Torr. After filling the helium bath of the cryostat with liquid $^4$He the $N_2$ gas present in the tube of the vacuum chamber of the calorimeter at $P$~$10^{-5}$Torr condensed onto the tube walls. The mass of condensed nitrogen was $10^{-9}$ mol Thus, practically all the nitrogen available in the vacuum chamber of the calorimeter was adsorbed by the c-SWNT bundles.

## 3. Results and discussion

The experimental temperature dependence of the total heat capacity $C_{ad+N2}$ and its "addenda" part $C_{ad}$ are shown in Fig. 2 over the temperature ranges 2-40 K (Fig. 2a) and 2-6 K (Fig. 2b), respectively. As can be seen, saturation of c-SWNT bundles with $N_2$ to the occupancy $\xi_{N2}$ ≈0.0066 caused a significant increase in the heat capacity over the entire temperature range. The ratio $C_{ad+N2}/C_{ad}$ is about 1.6 at 2 K < $T$ < 15 K and decreases to ~1.2 at 40 K.

The contribution $C_{N2}$ of nitrogen to the total heat capacity $C_{ad+N2}$ was separated by subtracting $C_{ad}$ from $C_{ad+N2}$. The heat capacity $C_{N2}$ was estimated assuming that the contributions $C_{ad}$ and $C_{N2}$ are additive because the influence of the 1D chains of $N_2$ molecules on the phonon density of the c-SWNT bundles is negligible. The coupling between the acoustic vibrations of the adsorbate atoms (molecules) in the chain and the carbon atoms in the nanotubes influences the physical properties of the system at much lower temperatures than in our experiment (see [48,64,65]).

Fig. 3 shows the normalized experimental heat capacity ($C_{N2}/(\mu R)$, where $\mu$ is the number of the nitrogen moles and $R$ is the gas constant) of the 1D chains of $N_2$ molecules adsorbed in grooves of c-SWNT bundles at $T$ = 2 - 40 K (Fig 3a) and 2 - 8 K (Fig. 3b). For comparison, Fig. 3 contains a theoretical curve of the molar heat capacity $(C_V/R)_{Kr}$ of the 1D chains of Kr atoms adsorbed in grooves of c-SWNT bundles [47]. It is seen (Fig. 3b) that the experimental heat capacity of nitrogen $C_{N2}/(\mu R)$ and the theoretical phonon heat capacity of krypton $(C_V/R)_{Kr}$ [47] are close within the range $T = 2 - 8$ K. Thus, below 8 K the heat capacity of the one-dimensional nitrogen chains in the grooves is influenced predominantly by the phonon modes.

The phonon heat capacity $C_V$ of physically adsorbed 1D chains of Xe, Kr, Ar, Ne atoms [47, 48] and CH$_4$ molecules [48] in grooves is determined by the longitudinal acoustic $L$ and two transverse optical $T_1, T_2$ modes. The contribution of the optical modes decreases exponentially with lowering temperature. At low temperatures, at which the contribution of the $L$ mode is dominant, the specific heat $C_V(T)$ can be presented in terms of the Debye model [47]:

$$\frac{C_V}{R} \approx 2.095 \cdot \frac{k_B T}{\hbar \omega_L (q = \pi/a)} = 2.095 \cdot \frac{T}{\Theta_{D,L}}, \quad (1)$$

where $k_B$ is the Boltzmann constant, $\hbar$ is the Plank constant, $\omega_L(q=\pi/a)$ is the highest Brilouin zone edge frequency of the $L$-mode, $a$ is the distance between the adsorbate particles in the chain. The Debye temperature $\theta_{D,L}$ is found using the sound velocity $\upsilon_L$ [48]

$$\theta_{D,L} = \pi \hbar \upsilon_L / a k_B. \quad (2)$$

Eq. (1) neglects all contributions other than longitudinal mode, limiting its applicability to an even lower temperature regime than usual $T << \theta_D$ [48, 64,65].

The solid lines in Fig. 3b demonstrate the linear behavior of the specific heat $C_{N2}/(\mu R)$ of 1D chains of N$_2$ molecules (straight line 1) and Kr atoms $(C_v/R)_{Kr}$ (straight line 2) [47] below 4 K. Using Eq (1), we obtain $\theta_{D,L} \approx 60$ K for the N$_2$ chains and the highest Brillouin zone edge frequency of the $L$ mode $\hbar \omega_L = 5.2$ meV.

Although the theoretical models [47,48] include significant simplifications, the experimental results $C_{N2}/(\mu R)$ and the theoretical predictions for the phonon heat capacity of krypton at $T < 8$ K are in good agreement.

Above 8 K the temperature dependence of the heat capacity of N$_2$ molecules $C_{N2}(T)/(\mu R)$ is steeper than that of Kr atoms $(C_V(T)/R)_{Kr}$ [47] (see Fig. 3). At higher temperatures, the difference $\Delta C = C_{N2}(T)/(\mu R) - (C_V(T)/R)_{Kr}$ between the heat capacities of the N$_2$ and Kr chains increases monotonically. It is reasonable to assume that above 8 K the difference $\Delta C$ is due to a contribution of orientational vibrations of the N$_2$ molecules (librons).

The temperature dependence of the heat capacity of $N_2$ chains (circles) and the experimental [59,60] (triangles) and theoretical [47] (solid curve) heat capacities of Xe chains are compared in Fig 4. Note that the experimental data for 1D chains of $N_2$ and Xe have been obtained in the same adiabatic calorimeter [60] and with the same sample of c-SWNT bundles [6]. The dependences are qualitatively similar at $T < 12$ K.

In the case of 1D Xe chains the experimental curve $C_{Xe}(T)$ goes above the theoretical one for $T > 8$ K and the excess $(\Delta C_{ph})$ increases steadily with temperature going up (see Fig.4). It was assumed [60] that $\Delta C_{ph}$ is mainly due to an increasing distance $a_{Xe}$ between nearest neighbor Xe atoms in the chain at higher temperatures. The theoretical curve for Xe chains was calculated assuming $a_{Xe}$ = const. The effect of the changes in $a_{N2}$ with increasing temperature upon the phonon heat capacity should also be observed for 1D chains of $N_2$ molecules above 8 K.

Thus, above 8 K the heat capacity of the 1D chains of $N_2$ in the grooves is determined by the translational vibrations of the centers of gravity of the molecules, the orientational vibrations of molecules as a whole and by the $\Delta C_{ph}$ variations when the distance $a_{N2}$ between the nearest neighbor $N_2$ molecules increases as temperature goes up.

**Conclusions**

The heat capacity of c-SWNT bundles containing 1D chains of adsorbed $N_2$ molecules in the grooves of the c-SWNT bundles has been investigated for the first time. The contribution of $N_2$ molecules $C_{N2}$ to the total heat capacity has been separated. The highest Brillouin zone edge frequency of the longitudinal phonon mode has been estimated for 1D chains of $N_2$ molecules in the grooves ($\hbar\omega_L = 5.2\ meV$). The contribution of the translational vibrations of $N_2$ molecules to the heat capacity of nitrogen is dominant at $T = 2 - 8$ K. Above 8 K the contribution of the orientational vibrations of $N_2$ molecules becomes significant.

**Acknowledgement**

The authors are indebted to V.G. Manzhelii, M.A. Strzhemechny, K.A. Chishko, S.B. Feodosyev, E.S. Syrkin, E.V. Manzhelii and I.A. Gospodarev for helpful discussions.

Fig.1. Possible sites of adsorption of relatively small impurity atoms or molecules in a c-SWNT bundle.

Fig. 2 a,b. Experimental temperature dependence of the total heat capacity $C_{ad+N2}$ (open circles) and its "addenda" part $C_{ad}$ (solid circles) in the temperature ranges 2-40 K (a) and 2-6 K (b).

Fig. 3. The normalized experimental heat capacity $C_{N2}(T)/(\mu R)$ of 1D chains of $N_2$ molecules adsorbed in the grooves of c-SWNT bundles at $T = 2 – 40$ K (Fig. 3a) and $T = 2 – 14$ K (Fig. 3b). The solid curve is the theoretical heat capacity $(C_V/R)_{Kr}$ of phonon modes (longitudinal acoustic $L$ mode and two transverse optical $T_1$ and $T_2$ modes) of adsorbed 1D chains of Kr atoms [47] in the grooves of c-SWNT bundles. The solid straight lines (Fig. 3b) show a linear low temperature behavior of the heat capacity of the longitudinal mode $L$ of the 1D chains of $N_2$ molecules (1) and Kr atoms (2).

Fig.4. The molar heat capacity of 1D adsorbate chains in the grooves. Experiment: $N_2$ – circles (this work); Xe –triangles [59,60]. Theory: Xe– solid curve [47].

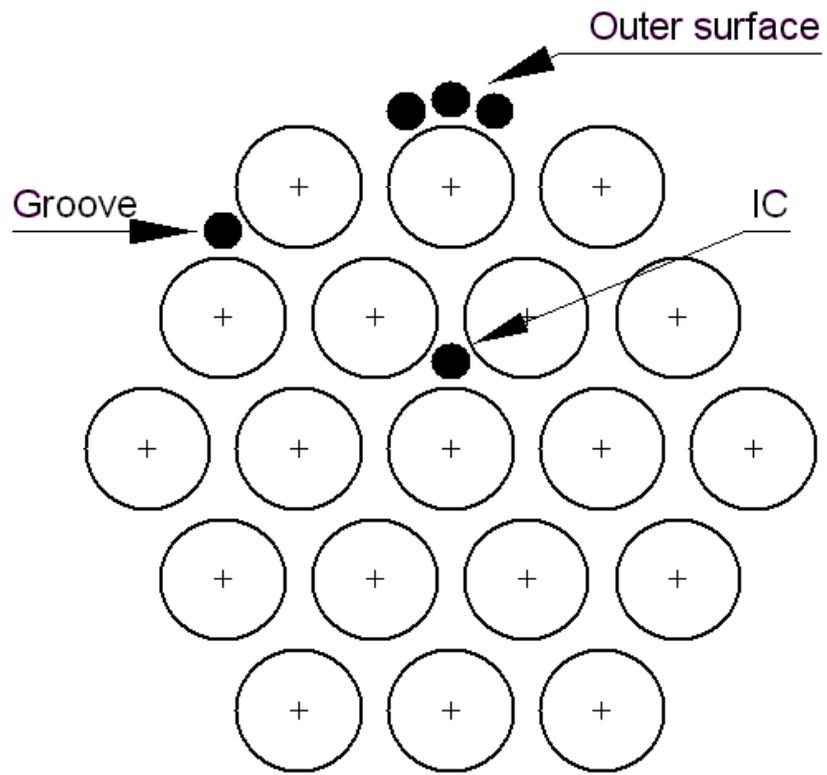

Fig. 1.

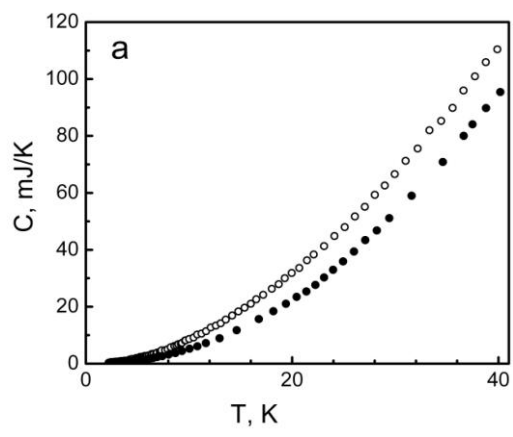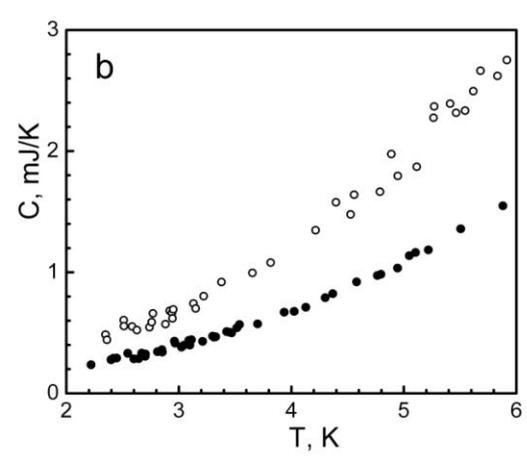

Fig. 2

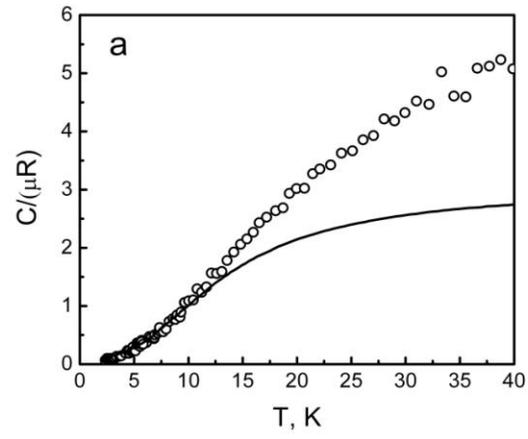

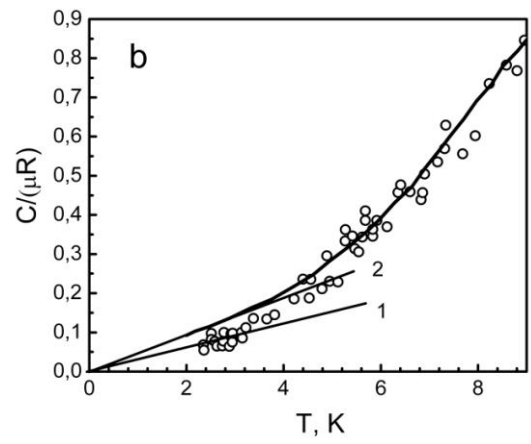

Fig. 3.

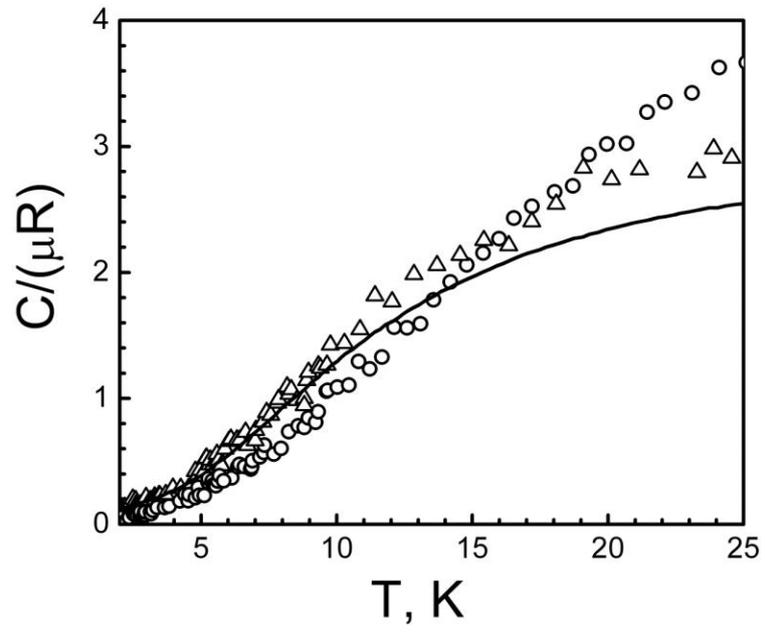

Fig. 4.